# Tracking Performance of Incremental LMS Algorithm over Adaptive Distributed Sensor Networks

Ehsan Mostafapour, C. Ghobadi, Javad Nourinia and M. Chehel Amirani
*Department of electrical and computer engineering*
*Urmia University, Urmia-Iran*
{E.mostafapour, ch.ghobadi, j.nourinia, m.amirani} @urmia.ac.ir
Corresponding author Email:E.mostafapour@urmia.ac.ir

*Abstract*—In this paper we focus on the tracking performance of incremental adaptive LMS algorithm in an adaptive network. For this reason we consider the unknown weight vector to be a time varying sequence. First we analyze the performance of network in tracking a time varying weight vector and then we explain the estimation of Rayleigh fading channel through a random walk model. Closed-form relations are derived for mean square error (MSE), mean square deviation (MSD) and excess mean square error (EMSE) of analyzed network in tracking Rayleigh fading channel and random walk model. Comparison between theoretical and simulation results shows a perfect match and verifies performed calculations.

*Index Terms*—Distributed estimation, incremental strategies, non-stationary environment, Rayleigh fading.

I. INTRODUCTION

Wireless sensor networks are useful tools for a variety of tasks from environment monitoring to spectrum sensing [1-7].

Up until now there has been an excessive research on the topic of the performance of adaptive wireless sensor networks in different environmental conditions like extra noise addition, link failures and fading effects. All kinds of transient and steady-state behaviors of different consensus strategies have been analyzed and proposed in several papers [1, 2], but in all of these papers the main assumption about the usage of network is that the network is designed to estimate a stationary entity. So what if we want to use the network to track a non-stationary variable? As stated in [1], consensus strategies are more capable in improving the performance of adaptive estimation than any other variations. In short we can understand that group estimation has a better result than singular case. This ability can be used in tracking a non-stationary entity. In this paper we address this issue and propose some usages for the tracking with an adaptive wireless sensor network.

Fading channel estimation and object tracking can be the main examples of tracking a non-stationary variable. Here we considered the first application namely fading channel tracking. The reader should be aware that this criterion is different from estimating a stationary variable in a fading channel. Here our goal is to track a non-stationary variable in a stationary environment. This environment is subject as usual to white Gaussian noise but not fading. Even we can expand this case to the case where our environment is subjected to fading itself. Here we will overview the topic of discussion in some recent papers:

In [4] the tracking performance is mentioned and some attempt is made to reach closed-form expressions for MSD of a diffusion strategy in non-stationary environment, but in simulation part no relations is mentioned about the relevance of simulated scenario and calculated theoretical results. Furthermore no matching diagram is given to support the arrived simulation and theoretical results which is a verification of the accuracy of results.

In [5] following the scheme of [4], tracking performance of a variable step-size diffusion LMS algorithm is considered in non-stationary environment, but this time no attempt is made to arrive to a closed-form expression of MSD or MSE of the network and consequently no matching is performed in between simulation and theoretical results.

The authors of [6] assumed that the fading coefficients of the network are known to us and therefore we can adjust combination parameters in a way to mitigate deep fading transitions. This is true when you estimate fading channel coefficients in advance and for this problem we can use our network to track the changing fading conditions. Also in [6] no closed-form relations are given for this scenario.

Finally in reference [3] theoretical results are given for performance of Distributed Incremental LMS (DILMS) algorithm in a non-stationary environment. But in simulation part the performance result of IDLMS algorithm is presented in tracking a time varying auto regressive (TVAR) sequence and consequently the results are matched without a relevance to Random-walk model.

In this paper we match the Rayleigh channel estimation problem with Random-walk model for non-stationary environments and consider several models for changing and drifting unknown weights. In this situation we compare the MSD and EMSE of simulations with theoretical results directly.

This paper consists of the following parts: in part II we will have a brief overview of incremental adaptive estimation of a stationary data in distributed sensor networks and then we will start to develop non-stationary data model. In part III we study the tracking performance of IDLMS algorithm and overview the derivation of theoretical Steady-state performance results. In part IV we present our simulation examples for tracking of time varying weights and Rayleigh fading channel estimation. Part V contains our concluding remarks and suggestions for future works.

TABLE I. USED SYMBOLS AND THEIR DESCRIPTION

| Symbol | description |
|---|---|
| $(.)^T$ | Transposition |
| $\mathbb{E}(a)$ | Statistical expectation of $a$ |
| $(.)^*$ | Conjugation for scalars and Hermitian transpose for matrixes |
| $\|x\|_\Sigma^2$ | $x^*\Sigma x$ for a column vector x |
| Tr[A] | Trace of matrix A |

**Notation:** In this paper as in [1] and [2], we used bold faced letters for random quantities and plain text letters for deterministic quantities. Upper case letters are also used for matrixes. Symbols which are used in this paper are explained in TABLE I:

## II. DISTRIBUTED ESTIMATION

Following the procedure in [1], we will consider a network of *N* nodes distributed in an area (Fig.1). If we take k as sensor index, each node has access to time realizations $\{d_k(i), u_{k,i}\}$ of $\{\boldsymbol{d}_k(i), \boldsymbol{u}_{k,i}\}$ measurements. For each sensor $\boldsymbol{d}_k$ is a scalar quantity and, $\boldsymbol{u}_k$ is a $1 \times M$ vector. The data model for stationary case represents the relation between these measurements through a linear equation [1]:

$$\boldsymbol{d}_k(i) = \boldsymbol{u}_{k,i} w^o + \boldsymbol{v}_k(i) \tag{1}$$

Where $w^o$ is the desired unknown vector and $\boldsymbol{v}_k(i)$ is white noise with variance $\sigma_{v,k}^2$. Same independence assumptions are taken into consideration as in [1]. In stationary case the objective of the adaptive network is to estimate desired deterministic vector $w^o$ through minimizing the mean square error [3]:

$$w^o = \underset{w}{\arg\min} \frac{1}{N} \sum_{k=1}^{N} \mathbb{E}|\boldsymbol{d}_k - \boldsymbol{u}_k w|^2 \tag{2}$$

The optimal weight is then [2]:

$$w^o = \left(\sum_{k=1}^{N} R_{u,k}\right)^{-1} \left(\sum_{k=1}^{N} R_{du,k}\right) \tag{3}$$

Where $R_{u,k} = \mathbb{E}(\boldsymbol{u}_k^* \boldsymbol{u}_k)$ and $R_{du,k} = \mathbb{E}(\boldsymbol{d}_k \boldsymbol{u}_k^*)$ are correlation terms.

### A. Distributed Incremental LMS (DILMS) estimation

As mentioned before based on the cooperation strategy between nodes we have two main choices: Incremental and Diffusion strategies. In this paper we will analyze the tracking performance in the incremental mode. In incremental strategy each node only has communication with its immediate neighbors [1]. This makes Incremental strategy more cost efficient based on computational

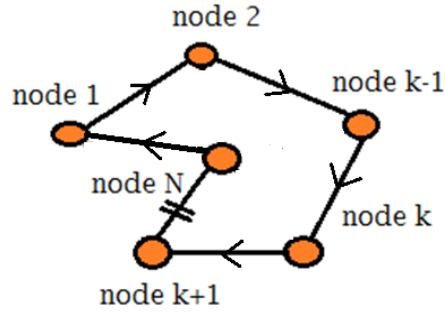

*Fig.1.* A distributed network with (*N*) active nodes and incremental mode of cooperation

complexity.

The complexity study is never considered fully in distributed adaptive strategies. Although we can give precise calculation amount for adaptive algorithms, when it comes to the collaborative work of 20 or more sensor nodes, the tracking of computational cost is a formidable task. In reference [1] it is mentioned that for IDLMS algorithm, if the length of unknown vector is taken to be ($M$) we need $O(M)$ (order of M) computations per node and $O(M)$ transmissions per node. For our simulations $M = 4\ and\ 8$. It is also claimed in [1] that for other similar algorithms, for each node $O(M^3)$ computational complexity and $O(M^2)$ transmission is needed. For example in diffusion LMS algorithm, the transmission between nodes is more than IDLMS algorithm because each node has communications with more than one neighboring node.

In this strategy each node shares local estimation of desired vector namely $\boldsymbol{\psi}_{k,i}$ with its immediate neighbor and plays as the initial estimation for the next node. The IDLMS algorithm is given as [1]:

For each time $i \geq 0$ repeat:

$$\begin{cases} \boldsymbol{\psi}_{0,i} = \boldsymbol{w}_{i-1} \text{initial guess} \\ \boldsymbol{\psi}_{k,i} = \boldsymbol{\psi}_{k-1,i} - \mu_k u_{k,i}^*(d_k(i) - u_{k,i}\psi_{k-1,i}),\ k = 1, \dots, n \\ \boldsymbol{w}_i = \boldsymbol{\psi}_{n,i} \text{ at node } N \end{cases} \quad (4)$$

*B. Non-stationary environment*

Different papers proposed different modes for non-stationary environments. In [4] and [5] the proposed time varying vector of length 8 (or $M = 8$) is defined as:

$$\boldsymbol{w}_i^o = \tfrac{1}{2}[a_{1,i}, a_{2,i}, a_{3,i}, a_{4,i}]^T \quad (5)$$

Where $a_{k,i} = \left[cos\left(\omega i + \frac{(k-1)\pi}{2}\right), sin\left(\omega i + \frac{(k-1)\pi}{2}\right)\right]$ for $k = 1,2,3,4$ and $\omega = \frac{\pi}{3000}$. Also in [3] a time varying auto regressive (TVAR) model is proposed for defining non-stationary environment.

In this paper we propose a much realistic mode for non-stationary environment. Here we assume that our unknown weight vector takes root from a Rayleigh fading channel. In this case the objective of the sensor network is to track and estimate the time varying channel coefficients. Rayleigh fading channel assumption can be changed by other channel modes but due to the simplicity of calculations and following the procedure in [8] we have the channel coefficients as:

$$h(n) = \gamma x(n)\delta(n - n_0) \tag{6}$$

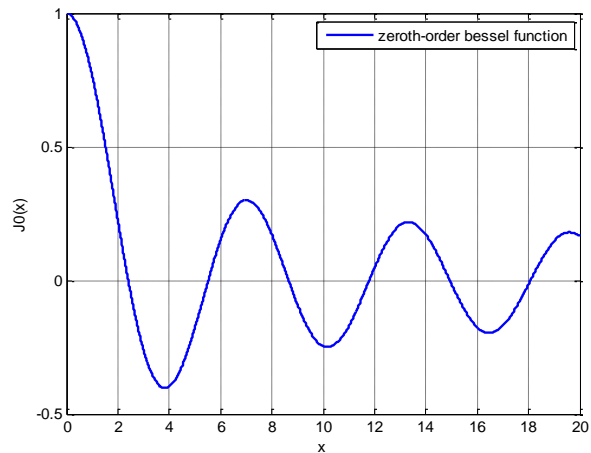

*Fig.2.* Zero-order Bessel function

Where $\gamma$ is path loss and $n_0$ is the channel delay. $x(n)$ is a time varying sequence and its amplitude $|x(n)|$ is assumed to have Rayleigh distribution as follows[7]:

$$f_{|x(n)|}(|x(n)|) = |x(n)|e^{-|x(n)|^2/2}, \quad |x(n)| \geq 0 \tag{7}$$

The phase of this distribution is uniformly distributed within $[-\pi, \pi]$. The auto-correlation of the sequence $x(n)$ is modeled as zeroth-order Bessel function:

$$r(k) = \mathbb{E}[x(n)x(n-k)] = \mathcal{J}_0(2\pi f_D T_s k), \quad k = \cdots, -1, 0, 1, \ldots \tag{8}$$

Where $f_D$ is the Doppler frequency, $T_s$ is the sampling period and $\mathcal{J}_0(.)$ is the Bessel function that is shown in *Fig*. 2. A first order approximation for the variation of a Rayleigh fading coefficient $x(n)$ is to assume it varies according to an auto regressive model [8]:

$$x(n) = r(1)x(n-1) + \sqrt{1 - |r(1)|^2}\, v(n) \tag{9}$$

Where $r(1) = \mathcal{J}_0(2\pi f_D T_s)$ and $v(n)$ is a white noise process with unit variance.

Now we can assume a weight vector with the entries of these Rayleigh fading coefficients and as the coefficients change at the same rate the above approximation indicates that the variations in weight vector could be approximated as:

$$\boldsymbol{w}_i^o = \alpha \boldsymbol{w}_{i-1}^o + \boldsymbol{\eta}_i \tag{10}$$

Where $\boldsymbol{\eta}_i$ is the non-stationary variable part of random-walk model with the covariance matrix [7]:

$$R_\eta = (1 - \alpha^2)I \tag{11}$$

With $\alpha = r(1)$ and $I$ to be a unit matrix. The value of $\alpha$ depends on Doppler frequency. This how we produce the time varying weight vector and in other words we change the unkown vector estimation problem to a channel estimation problem with the help of WSNs. Further explanations are given in simulation part.

III. TRACKING PERFORMANCE

In this section we will present the necessary calculations towards mean-square performance of IDLMS algorithm in tracking task. Then we will apply these calculations to the proposed Rayleigh fading channel estimation problem.

First we start with our data model for tracking performance:

$$\boldsymbol{d}_k(i) = \boldsymbol{u}_{k,i}\boldsymbol{w}_i^o + \boldsymbol{v}_k(i) \tag{12}$$

Where $\boldsymbol{w}_i^o$ is given in (10), $\boldsymbol{u}_{k,i}$ is the regressor vector and $\boldsymbol{v}_k(i)$ is additive noise. Now we define weight error vector:

$$\widetilde{\boldsymbol{w}}_i \triangleq \boldsymbol{w}_i^o - \boldsymbol{w}_i \tag{13}$$

The mean of this vector can be written as [3]:

$$\mathbb{E}\widetilde{\boldsymbol{w}}_i = \left[1 - \mu \sum_{k=1}^{N} R_{u,k}\right] \mathbb{E}\{\widetilde{\boldsymbol{w}}_{i-1} + \boldsymbol{\eta}_i\} \tag{14}$$

In (14), $\boldsymbol{\eta}_i$ is a zero mean variable sequence with covariance matrix $R_\eta$. Our purpose is to achieve mean square deviation (MSD) and excess mean square error (EMSE) for each node defined as:

$$MSD_k = \mathbb{E}\|\widetilde{\boldsymbol{w}}_{k,\infty}\|_I^2 \tag{15}$$

$$EMSE_k = \mathbb{E}\|\widetilde{\boldsymbol{w}}_{k,\infty}\|_{R_{u,k}}^2 \tag{16}$$

For the IDLMS algorithm in (4), the error signals can be defined as follows:

$$\widetilde{\boldsymbol{\psi}}_{k,i} \triangleq \boldsymbol{w}_i^o - \boldsymbol{\psi}_{k,i} \tag{17}$$

$$\boldsymbol{e}_k(i) \triangleq \boldsymbol{d}_{k,i} - \boldsymbol{u}_{k,i}\widetilde{\boldsymbol{\psi}}_{k-1,i} \tag{18}$$

The variance relation for IDLMS algorithm in stationary case can be used to express mean square behavior of this algorithm:

$$\mathbb{E}\|\overline{\boldsymbol{\psi}}_{k,i}\|_{\overline{\sigma}_k}^2 = \mathbb{E}\|\overline{\boldsymbol{\psi}}_{k,i-1}\|_{\Pi_{k+1,1}\overline{\sigma}_k}^2 + a_{k+1}\overline{\sigma}_k \tag{19}$$

For Gaussian regressor data with Eigen decomposition $R_{u,k} = U_k \Lambda_k U_k^*$, matrix $\Pi_{k,l}$ is defined as [1]:

$$\Pi_{k,l} = \bar{F}_{k+l-1}\bar{F}_{k+l} \dots \bar{F}_N \bar{F}_1 \dots \bar{F}_{k-1} \tag{20}$$

Where

$$\bar{F}_k = I - 2\mu_k \Lambda_k + 2\mu_k^2 \Lambda_k^2 + \mu_k^2 \lambda_k \lambda_k^T \tag{21}$$

In this equation $\lambda_k$ is a vector containing diagonal enteries of $\Lambda_k$.

In equation (19) with the definition $g_k = \mu_k^2 \sigma_{v,k}^2 \lambda_k^T$, we can define row vector $a_k$ as [1]:

$$a_k = g_k \Pi_{k,2} + g_{k+1}\Pi_{k,3} \dots + g_{k-2}\Pi_{k,N} + g_{k-1} \tag{22}$$

And $\bar{\sigma}_k$ can be achieved with the transformed form of a non-negative matrix $\Sigma$ [2]:

$$\bar{\Sigma} = U_k^* \Sigma U_k, \quad \bar{\sigma}_k = diag(\bar{\Sigma}) \tag{23}$$

Now if we define:

$$\bar{\boldsymbol{\psi}}_k = U_k^* \tilde{\boldsymbol{\psi}}_k \tag{24}$$

Then equation (19) for a non-stationary condition with $\boldsymbol{w}_i^o = \alpha \boldsymbol{w}_{i-1}^o + \boldsymbol{\eta}_i$ can be written as:

$$\mathbb{E}\left\| U_k^*(\boldsymbol{w}_i^o - \boldsymbol{\psi}_{k,i}) \right\|_{\bar{\sigma}_k}^2 = \mathbb{E}\left\| U_k^*(\boldsymbol{w}_{i-1}^o + \boldsymbol{\eta}_i - \boldsymbol{\psi}_{k,i-1}) \right\|_{\Pi_{k+1,1}\bar{\sigma}_k}^2 + a_{k+1}\bar{\sigma}_k \tag{25}$$

Using (17) we can write:

$$\mathbb{E}\left\| U_k^* \tilde{\boldsymbol{\psi}}_{k,i} \right\|_{\bar{\sigma}_k}^2 = \mathbb{E}\left\| U_k^*(\tilde{\boldsymbol{\psi}}_{k,i-1} + \boldsymbol{\eta}_i) \right\|_{\Pi_{k+1,1}\bar{\sigma}_k}^2 + a_{k+1}\bar{\sigma}_k \tag{26}$$

With the definition of $\bar{\boldsymbol{\eta}}_i = U_k^* \boldsymbol{\eta}_i$, the variance relation for IDLMS algorithm for the non stationary condition of (10) is given as [3]:

$$\mathbb{E}\left\| \bar{\boldsymbol{\psi}}_{k,i} \right\|_{\bar{\sigma}_k}^2 = \mathbb{E}\left\| \bar{\boldsymbol{\psi}}_{k,i-1} \right\|_{\Pi_{k+1,1}\bar{\sigma}_k}^2 + \mathbb{E}\|\bar{\boldsymbol{\eta}}_i\|_{\Pi_{k+1,1}\bar{\sigma}_k}^2 + a_{k+1}\bar{\sigma}_k \tag{27}$$

In this relation a diag(.) operator which is used for diagonalization is omitted, and we can write:

$$\mathbb{E}\|\bar{\boldsymbol{\eta}}_i\|_{\Pi_{k+1,1}\bar{\sigma}_k}^2 = Tr\left(\mathbb{E}\{\bar{\boldsymbol{\eta}}_i^* \bar{\boldsymbol{\eta}}_i\} diag(\Pi_{k+1,1}\bar{\sigma}_k)\right) = Tr\left(\bar{R}_\eta diag(\Pi_{k+1,1}\bar{\sigma}_k)\right) \tag{28}$$

Now we can write (27) as:

$$\mathbb{E}\left\| \bar{\boldsymbol{\psi}}_{k,i} \right\|_{\bar{\sigma}_k}^2 = \mathbb{E}\left\| \bar{\boldsymbol{\psi}}_{k,i-1} \right\|_{\Pi_{k+1,1}\bar{\sigma}_k}^2 + Tr\left(\bar{R}_\eta diag(\Pi_{k+1,1}\bar{\sigma}_k)\right) + a_{k+1}\bar{\sigma}_k \tag{29}$$

Steady-state performance of IDLMS algorithm can be obtained when the $i$ index goes to infinity:

$$\mathbb{E}\left\| \bar{\boldsymbol{\psi}}_{k,\infty} \right\|_{(I-\Pi_{k+1,1})\bar{\sigma}_k}^2 = Tr\left(\bar{R}_\eta diag(\Pi_{k+1,1}\bar{\sigma}_k)\right) + a_{k+1}\bar{\sigma}_k \tag{30}$$

The steady-state MSD of IDLMS algorithm for each node can be obtained by using the replacement $(I - \Pi_{k+1,1})\bar{\sigma}_k = q = diag(I)$, as [3]:

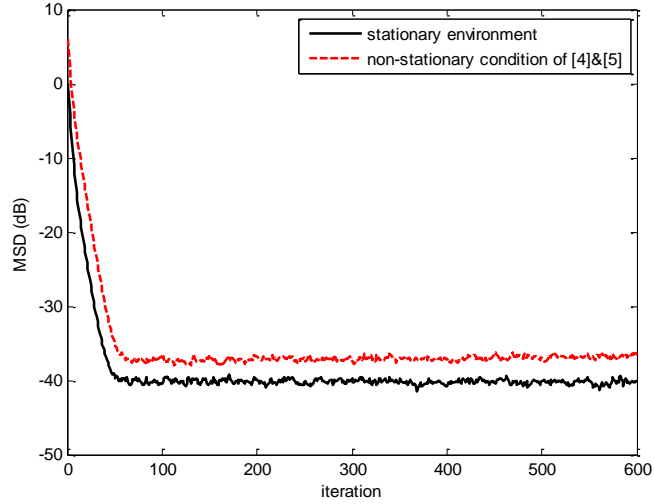

Fig.3. Performance comparison between time varying weight vector and stationary case

$$MSD_k = Tr\left(\bar{R}_\eta diag\left(\Pi_{k+1,1}(I - \Pi_{k+1,1})^{-1} q\right)\right) + a_{k+1}(I - \Pi_{k+1,1})^{-1} q \quad (31)$$

And steady-state EMSE can be given by replacing $\lambda_k = (I - \Pi_{k+1,1})\bar{\sigma}_k$ [3]:

$$EMSE_k = Tr\left(\bar{R}_\eta diag\left(\Pi_{k+1,1}(I - \Pi_{k+1,1})^{-1} \lambda_k\right)\right) + a_{k+1}(I - \Pi_{k+1,1})^{-1} \lambda_k \quad (32)$$

In both (31) and (32) $\bar{R}_\eta$ is defined as:

$$\bar{R}_\eta = \mathbb{E}\{\bar{\boldsymbol{\eta}}_i^* \bar{\boldsymbol{\eta}}_i\} \quad (33)$$

## IV. SIMULATION RESULTS

Here we explain and present our simulation results for tracking performance of adaptive incremental algorithms in distributed sensor networks. Two scenarios are taken to the consideration. First we adopt the time varying weight vector of [4] and [5] and then we will proceed to our proposed channel estimation problem and match theoretical and simulation results. In all of our simulations we assume a network with 20 nodes. It is important to mention thatby increasing the number of nodes, the number of iterations for convergence decreases but the amount of computation increases heavily. This is obvious from the simulation running time. All simulation results are averaged over 60 Monte Carlo runs. The steady-state curves are obtained by averaging the last 500 iteration results of simulations [2]. The variance of noise is considered to be the same for all nodes and we have $\sigma_{v,k}^2 = 0.01$. Also the step-size is considered to be fixed and it equals to 0.0045.

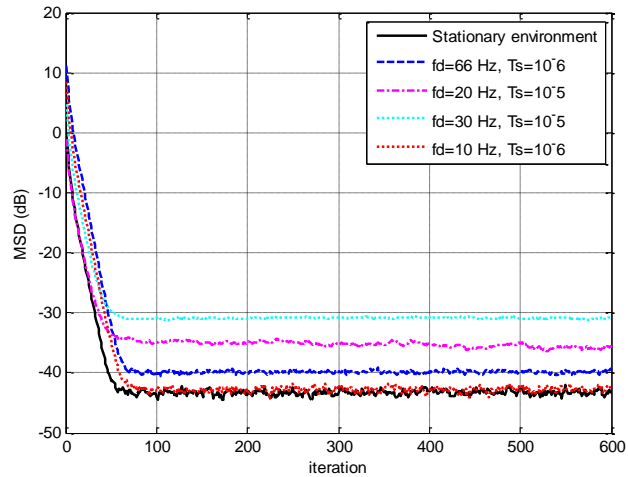

Fig.4. performance comparison in channel estimation

### A. Time varying weight vector

Here we assume that our unknown weight vector is changing according to (5). In this case our weight vector is assumed to have eight entries. The dynamic model for $w_i^o$ is [4]:

$$w_i^o = e^{j\omega} w_{i-1}^o \tag{34}$$

For changing this model to random walk model we can use mathematical operations but here we only present simulation results for time varying vector model and theoretical results are only provided for channel estimation scenario. In these conditions the tracking MSD of IDLMS algorithm is given in *Fig*. 3. The results are compared with fixed weight vector scenario (Stationary environment). It is obvious that the performance of algorithm in non-stationary environment is degraded comparing to stationary case. Similar results for diffusion strategy can be found in [4].

### B. Rayleigh fading channel estimation

In this scenario we consider a weight vector with 4 entries ($M = 4$) that are drawn from the model explained in (9) and (10). The MSD performance of IDLMS algorithm in estimating Rayleigh fading channel coefficients is given in *Fig*.4 for different Doppler frequency and sampling period quantities. In [15] it is mentioned for operating frequencies between 100 MHz and 2 GHz Doppler frequency shift can be as large as 128 Hz. Also in [8] sampling period considered between $10^{-6}$ to $10^{-3}$ seconds. In this case the diagonal entries of covariance matrix in (11) can be derived using (8). With these assumptions we have channel coefficient estimation results.

It is important to mention that when $T_s = 10^{-6}$ and $f_D = 10$ we have $\alpha = r(1) = \mathcal{J}_0(2\pi f_D T_s) \approx 1$ and so the MSD performance of channel estimation is almost similar to stationary senario. It is

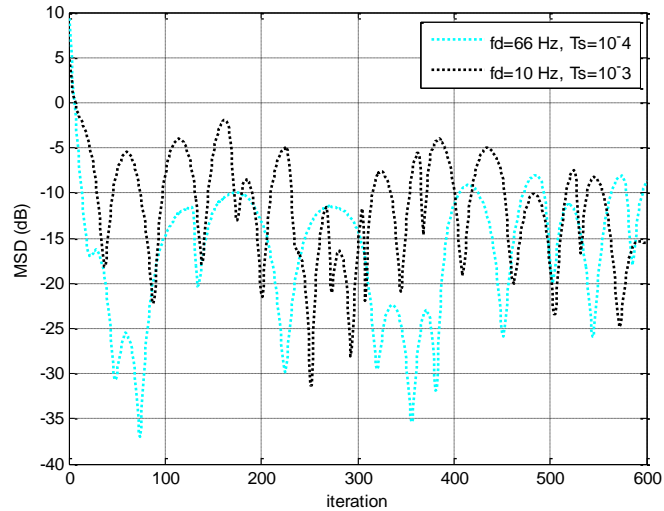

*Fig.5* comparison between theoretical and simulation results

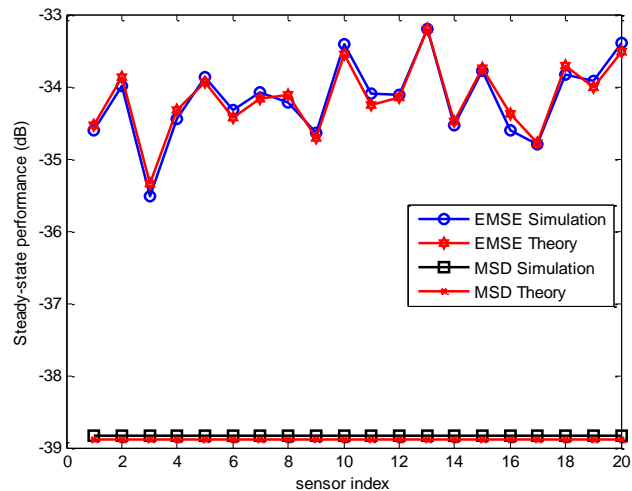

*Fig.6* comparison between theoretical and simulation results obvious from *Fig*. 5 that as Doppler frequency and sampling duration increase, the performance of network in tracking the changes in unknown vector degrades and when $T_s$ is more than $10^{-3}$, the algorithm may not converge. It is important to mention that the sampling period $T_s$ in these simulations is so low (order of $10^{-6}$ Seconds) that the changes of channel during these samples cannot be sensed visually. To address this issue we performed some simulations with longer sampling periods (order of $10^{-3}$ seconds) and concluded that in these conditions, the convergence diagram will ripple heavily. In Fig.5 we present some of these results.

Now, it is time to compare simulation results with our theoretical results. For this reason first we may pick a certain Doppler frequency and sampling period and using *Fig. 2* we have access to matrix $Q$. Then using equations (31) and (32) by invoking results in [12] we can find the exact amounts of EMSE and MSD. In *Fig*. 6 the steady state MSD and EMSE of IDLMS algorithm is presented and compared with theoretical results. For this simulation we assumed a fading channel with Doppler frequency of 66 Hz and sampling period of $10^{-6}$.

## VI. DISCUSSION AND RESULTS ANALYSIS

In this paper we simulated and analyzed the performance of distributed incremental LMS algorithm in tracking time varying vectors. Two main scenarios are taken into consideration. First we assumed that the unknown vector changes according to (5) and performed simulations for this situation and compared it with stationary data estimation. *Fig*. 3 showed that the performance degrades in non-stationary scenario and it was an expected result. For this simulation no theoretical analysis is performed. Next we assumed that we want to estimate a Rayleigh fading channel and unknown vector entries are channel coefficients. We simulated this assumption for different Doppler frequency and Sampling period amounts and concluded that as these two factors increase, the performance of the network degrades. The theoretical analysis for channel estimation scenario is then applied to our simulations and as in *Fig. 6* there is a reasonable match between theoretical and simulation results.

## VII. CONCLUSION

Over all in this paper we showed that the incremental distributed LMS algorithm can be used for tracking a non-stationary and time varying weight vector. The simulation results confirmed that as the variations of unknown vector in time gets more serious, the performance of network degrades. In future works we will perform channel estimation with distributed networks for a variety of fading channel types. Also the usage of more complicated distributed strategies like Diffusion strategy can help to improve performance results.